\documentclass{elsart}
\usepackage{natbib}
\usepackage{graphicx}

\def\gapprox{\lower.4ex\hbox{$\;\buildrel >\over{\scriptstyle\sim}\;$}}
\def\lapprox{\lower.4ex\hbox{$\;\buildrel <\over{\scriptstyle\sim}\;$}}
\def\bk{\mbox{\boldmath $k$}}

\def\be{\mbox{\boldmath $e$}}

\def\bB{\mbox{\boldmath $B$}}
\def\bb{\mbox{\boldmath $b$}}

\begin{document}

\begin{frontmatter}

\title{X-ray emission from magnetic dissipation in the magnetar magnetosphere
}

\author{Qinghuan Luo}

\address{School of Physics, The University of Sydney, NSW 2006, Australia}

\begin{abstract}
Magnetic dissipation through decay of Alfv\'en waves in the magnetar 
magnetosphere is discussed. Transport of magnetic fields in the star
leads to dissipation of the magnetic energy through either direct internal 
heating or transferring of the energy in waves that decay in the magnetar
magnetosphere. In the latter case, the Alfv\'en waves are excited by crust 
dislocations or elastic waves underneath the star's surface. It is suggested 
that these Alfv\'en waves can decay into ion sound waves which can be effectively 
damped leading to strong plasma heating. Hot plasmas expand producing transient X-rays.
\end{abstract}
                                                                                                     

\end{frontmatter}

\section{Introduction}

Soft gamma-ray repeaters (SGR) and anomalous X-ray pulsars (AXP) are 
believed to be magnetars, which are extremely strongly magnetized, slowly rotating 
neutron stars with a rotation period $P=6-8\,\rm s$ and a magnetic field well 
exceeding the quantum electrodynamic (QED) critical field, 
$B\gg B_c\approx 4.4\times10^9\,\rm T$ \citep{td95}. 
Both SGR and AXP emit persistent thermal X-rays modulated by 
magnetar's rotation, with a luminosity much higher than their 
spin-down power, suggesting that the emission is powered by the
magnetic energy not the rotation energy~\citep{td95}. SGRs emit bursts 
of soft gamma-rays (or X-rays) as well as persistent, pulsed X-rays \citep{wetal06}. 
Although the prominent feature of AXPs is their persistent X-rays, at least 
one of them has been detected to have  SGR-like burst emission~\citep{k04}. 
It is generally thought that radiation processes in 
magnetars are predominantly driven by magnetic field decay in the neutron star's 
interior or crustal region, either through direct internal heating as the result 
of a dissipation process or through damping of waves that are excited by 
the release of the magnetic stress~\citep{cetal04}. 
This latter mechanism is considered here since no efficient internal 
dissipation has been identified so far. An enormous magnetic stress
built up in the crust as the result of the magnetic transport processes in the star
causes crustal dislocations or disturbances (e.g. shear waves) that excite fast-mode 
and Alfv\'en waves transferring a significant fraction of the magnetic energy to the
magnetar magnetosphere \citep{betal89,td95}. Since low-frequency fast-mode waves
or Alfv\'en waves generated in a superstrongly magnetized plasma are subject to negligible 
damping, they must undergo a cascade through wave-wave interactions. 
\citet{hh99,hh05} discussed that fast-mode waves may steepen and breakdown leading to
dissipation of wave energy. We consider here that shear waves propagate in a horizontal direction
in the crust region and that mode coupling produces predominantly Alfv\'en waves. 

Cascades of Alfv\'en waves through three-wave interactions from smaller to
larger wave numbers were discussed extensively in the context of the ISM turbulence
\citep{i64,k65,smm83,bn01,getal02,lg03}. This idea was extended
to the case of a superstrong magnetic field and it was suggested
as an important channel for dissipation of Alfv\'en waves in the magnetar 
magnetosphere~\citep{betal89,tb98}. Despite the extensive 
discussions in both the contexts of the ISM and magnetar, the specific 
dissipation mechanism is not well understood. Three-wave interactions in the incompressible MHD 
approximation lead to anisotropic turbulence with the cascade in
parallel (to the magnetic field) wavenumber suppressed~\citep{smm83,bn01,getal02,lg03}, 
while wave damping through kinetic processes is possible only when a 
cascade to large parallel wavenumbers occurs.  Although three-wave 
interactions involving the fast mode allow cascades in both parallel 
and perpendicular wavenumbers, these processes are important only below a 
critical frequency \citep{c05}.

In this paper, decay of Alfv\'en waves through three-wave interactions
into damping ion sound waves is considered as a possible dissipation 
mechanism in the magnetar magnetosphere. Decay of an Alfv\'en wave into 
the ion sound mode and a backward propagating Alfv\'en wave was 
suggested by \cite{go63} and discussed widely in terms of parametric or modulational
instability with the application to solar flares or the solar wind \citep{g78,d78,m91}.
In general, when density fluctuations are allowed, a pump wave (not necessary 
in a natural mode) decays into two side-band plus a density modulation.
In the low $v_s/v_A\ll1$ limit, where $v_s$ is the ion sound speed and
$v_A$ is the Alfv\'en speed, one side-band is in the Alfv\'en mode and
the density fluctuations are in the ion sound mode. The other side-band
is not in a natural mode and thus, it is not important. This process reduces
to three-wave decay \citep{go63}, with the growth rate for the instability 
(for the side band) being the most effective in the low $v_s/v_A\ll1$ limit. 
However, since one has a typical 
$v_s/v_A\sim 0.1$ in solar flares and $\sim 0.5$ in the solar wind, 
its applicability is limited. Here it is suggested that such process
may be efficient in magnetars where we have $v_s/c\sim 10^{-3}$. 
Since the presence of a propagating ion sound wave requires nonequibrilium 
between electron temperature and ion temperature, a condition
which is generally not satisfied for magnetars, the sound mode can be
treated as a quasimode \citep{m86} and an Alfv\'en wave can decay 
into an Alfv\'en wave propagating in the opposite direction
and an ion sound wave that is instantaneously damped. 
This process leads to conversion of the energy in the incoming wave 
to thermal kinetic energy in the plasma. It is proposed here that
plasma heating through such process leads to X-ray emission.

In Sec. 2, theory for generation of Alfv\'en waves is outlined and
decay of Alfv\'en waves through three-wave interactions involving 
ion sound waves is discussed in Sec. 3. A model for thermal X-ray emission
due to dissipation of these waves in magnetars is discussed in Sec. 4.  

\section{Excitation of Alfv\'en waves}

Magnetic transport in the neutron star can lead to magnetic stress
being built up in the crust which causes fractures or dislocations in the crust. 
Such events generate disturbances propagating, for example as shear waves 
that can couple to the Alfv\'en wave in the magnetosphere \citep{betal89}. 
Although shear waves can couple to both the Alfv\'en mode and fast mode,
here we discuss only the former where the transmitted energy is assumed to be
mainly in the Alfv\'en mode. One may treat the star's surface as a sharp boundary, a 
reasonable approximation because the scale height of the star's atmosphere is about 
a few cm, much less than the relevant wavelength. The typical frequency of 
shear waves in the crust can be estimated from
$\omega\sim V_{sh}/l_d=
10^5(V_{sh}/10^6\,{\rm m}{\rm s}^{-1})(l_d/10\,{\rm m})^{-1}\,
{\rm s}^{-1}$, where $l_d$ is the typical length scale of the displacement of
crust movement, $V_{sh}=(\mu/\rho)^{1/2}$ is the shear velocity,
$\mu\approx 10^{29}\,{\rm J}\,{\rm m}^{-3}$ is the shear modulus, and 
$\rho\approx 10^{17}\,{\rm kg}\,{\rm m}^{-3}$ is the crust density.

\begin{figure}
\includegraphics[width=7.5cm]{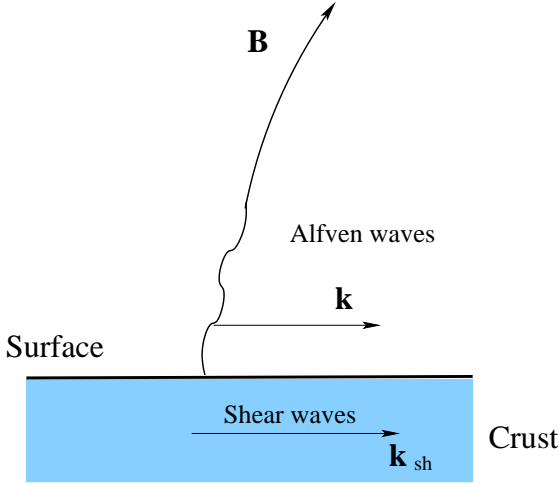}
\caption{Coupling of shear waves to the Alfv\'en mode. The surface is treated as
a sharp boundary so that the Snell's law applies: 
$k_{sh}\sin\alpha_{sh}\approx k\sin\alpha$.
Note that near the surface the propagation of the Alfv\'en mode
is mostly perpendicular, with $|k_\parallel|/k_\perp\sim V_{sh}/c\ll 1$.
}
\label{fig:alfv}
\end{figure}

When the surface is treated as a sharp boundary, Snell's law implies
that $k_{sh}\sin\alpha_{sh}=k\sin\alpha$, where
$k_{sh}$ and $\alpha_{sh}\approx \pi/2$ are wave number and incident angle (relative to the
surface norm) of the shear wave, and $k$ and $\alpha$ are the corresponding
wave number and incident angle of the transmitted Alfv\'en wave. 
Since the dispersion of the Alfv\'en mode can be written as
$\omega=|k_\parallel| v_0$, where $v_0=v_A/(1+v^2_A/c^2)^{1/2}$, 
the propagation angle can be determined from 
$\cos\theta=k_\parallel/k$. In the magnetar's strong magnetic field, 
one has the limit $v_A\gg c$ and this implies that $v_0\approx c$. 
Using $\omega\approx |k_\parallel| c\approx 
k_{sh}V_{sh}$, where the phase speed of the shear wave
is assumed to be $\sim V_{sh}$, one obtains $\cos\theta\approx V_{sh}/c$.
Thus, the transmitted Alfv\'en wave propagates nearly perpendicular
to the magnetic field at $\theta\approx\pi/2-V_{sh}/c$ (as shown in 
figure \ref{fig:alfv}). The mode coupling (to Alfv\'en waves) is efficient only in the surface
area where the magnetic field lines have an angle $\sim V_{sh}/c$ with 
respect to the surface normal direction.

A relevant issue here is how the Alfv\'en waves dissipate in the magnetosphere. 
The Landau damping rate is proportional to $\omega(\omega/\Omega_i)^2$, which 
is negligible for the frequency $\omega$ being much lower than the ion cyclotron 
frequency $\Omega_i$. In the absence of direct damping processes, the Alfv\'en waves
must undergo cascades through three-wave interactions. A widely discussed 
process is the three-wave interaction involving three Alfv\'en waves (denoted 
by $A$, $A'$, and $A'$): $A\to A'+A''$ or $A+A'\to A''$. Although a cascade in $k_\parallel$ is possible
in principle in the presence of a natural broadening in the three-wave
resonance, the cascade occurs predominantly in $\bk_\perp$ leading to
anisotropic Alfv\'enic turbulence \citep{smm83,bn01,getal02,lg03}. 
The Alfv\'en mode can be converted to a mixture of Alfv\'en and fast-mode 
($F$) waves through $A\to A'+F$.  Compared to pure Alfv\'enic turbulence, three-wave 
processes involving the fast mode are more efficient for frequency below a critical 
frequency $\omega_c$ (corresponding to a critical 
wavelength $v_0/\omega_c$) \citep{c05,lm06}.

\section{Decay into ion sound waves}

Alfv\'en waves can decay into ion sound waves through three wave interactions
$A\to A'+S$. The ion sound mode, like the normal sound wave, is longitudinal and
is heavily damped unless the ion temperature is much higher than the electron temperature.
The process is efficient in the low $\beta_s=v_s/v_0\ll 1$ limit, an approximation
relevant for the magnetar. Thus, this process provides an effective way to damp 
the Alfv\'en wave.

\subsection{Kinematic conditions}

\begin{figure}
\includegraphics[width=7.5cm]{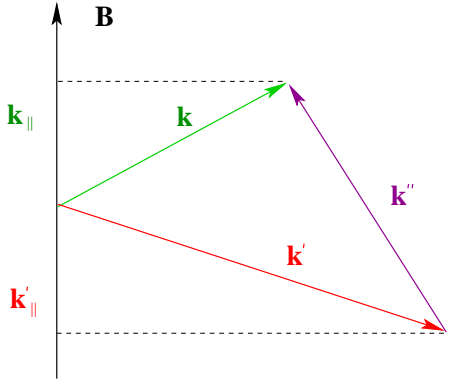}
\caption{Kinematic conditions of $A\to A'+S$. Since $v_A\gg v_s$ one has
$k_\parallel\approx -k'_\parallel$ and $k''_\parallel\approx 2k_\parallel$.}
\label{fig:wdecay}
\end{figure}

We consider the weak turbulence approximation in which the wave 
energy and momentum are conserved in each interaction. The energy 
and momentum conservation here, also referred to as the three-wave 
resonance condition, is described by $\omega=\omega'+\omega''$ and ${\bk}={\bk}'+{\bk}''$,
where $(\omega,{\bk})$, $(\omega',{\bk}')$ and $(\omega'',{\bk}'')$ represent
the frequencies and wave vectors of the incoming ($A$), scattered ($A'$) waves 
and the ion sound ($S$) wave, respectively. 
The condition on the wave vectors is shown in figure~\ref{fig:wdecay}.
The dispersion of the ion sound wave is $\omega''
=|k''_\parallel|v_s$ with $v_s=(\kappa_BT_e/m_p)^{1/2}
\approx 3\times10^5\, (T_e/10^7\,{\rm K})^{1/2}\,{\rm m}\,{\rm s}^{-1}$,
where $T_e$ is the electron temperature and we assume ions are protons. 
In the following we assume the incoming Alfv\'en wave propagates 
forward with respect to the magnetic field, $k_\parallel>0$.
The only permitted processes are those with $k'_\parallel<0$, corresponding
to an Alfv\'en wave decays into a sound wave and an oppositely directed 
Alfv\'en wave. From the resonance condition one has
$2k_\parallel=(1+\beta_s)k''_\parallel\approx k''_\parallel$ for
$\beta_s\sim v_s/c\ll1$. The decay rate can be determined by the growth rate of the 
backward propagating wave generated from the process. 
In the following we derive the growth rate in the broadband 
approximation where the relevant waves have a broad range of frequencies and wave vectors.

\subsection{Kinetic equations}

In the broadband approximation, a collection of waves in a 
particular mode can be described by the wave occupation number. Let 
$N_A(\bk)$, $N_{A'}(\bk')$ and $N_S(\bk'')$ be the occupation 
numbers of the incoming Alfv\'en wave, scattered Alfv\'en wave and 
sound wave. One can write down 
three kinetic equations that determine the rate of change of these three occupation 
numbers~\citep{m86}. To determine the efficiency of the decay we consider the evolution
of the scattered Alfv\'en wave. Since ion acoustic waves are heavily damped, one
can neglect $N_S$. For $\beta_s\ll1$, the kinetic equation for the scattered wave ($A'$) 
due to the processes $A\rightarrow A'+S$ and $A'\rightarrow A+S$ can
be written approximately as
\begin{eqnarray}
{d\ln N_{A'}(\bk')\over dt}&\approx&
\int{d\bk_\perp\over(2\pi)^2}
\,{\omega''\over v^2_0}\nonumber\\
&&\times{\partial\over\partial k_\parallel}\Bigl[
u_{AA'S}(\bk,\bk',\bk'')N_A(\bk)\Bigr]\Biggr|_{_{{\bf k}''={\bf k}-{\bf k}'}},
\label{eq:Gam1}
\end{eqnarray}
where $k_\parallel\approx -k'_\parallel$, $u_{AA'S}=\hbar
R_S|e_{i}e'_{j}{e''}^*_l\alpha_{ijl}|^2/(\varepsilon^3_0\omega\omega'\omega'')$ is
the probability for the transition $A\to A'+S$, $\alpha_{ijl}$ is the quadratic 
response tensor, $R_S$ is the ratio of the electric energy
to the total energy in the sound wave, and the corresponding ratio for the 
incoming and scattered waves is taken to be $R_A=R_{A'}=1/2$. The three vectors
$\be$, $\be'$ and $\be''$ are the polarizations of the three waves concerned.
The evolution of the scattered wave depends only the occupation number of
the incoming wave. 

\subsection{Growth rate}

The growth of the backward propagating Alfv\'en waves can be calculated 
from (\ref{eq:Gam1}). The three-wave probability can be derived 
assuming that electron motion in two high
frequency waves ($\omega$, $\omega'$) is treated in the cold plasma approximation.
Specifically, we assume the following conditions are satisfied: $\omega''/k''_\parallel=
v_s\ll v_e\ll \omega/k_\parallel =v_0$, where $v_e$ is the electron 
thermal speed. The quadratic response tensor has the following approximation~\citep{m86}
\begin{eqnarray}
\alpha_{ijs}&\approx& {e\varepsilon_0\omega''\over m_e}\tau_{ij}(\omega)k''_s\chi^{L(e)}(\omega'',\bk''),
\label{eq:a}
\\
\tau_{ij}&\approx& b_ib_j+{\rm i}{\omega\over\Omega_e}\varepsilon_{ijs}b_s,
\end{eqnarray}
where $\chi^{L(e)}\approx \omega^2_{pe}/{k''}^2v^2_e$ 
is the electronic contribution to the longitudinal component of the dielectric tensor,
$\bb=\bB/B$ is the magnetic field direction, $\omega_{pe}$ is the electron plasma
frequency, $\varepsilon_{ijs}$ is the permutation symbol (1 for even permutation,
$-1$ for odd permutation and 0 otherwise), and $\Omega_e$ is the electron cyclotron frequency.
The strong magnetic field approximation implies that only terms $\alpha_{123}=-\alpha_{213}$
are important. Since the component (\ref{eq:a}) corresponds to the order $1/\Omega_e$ terms in 
the quadratic response tensor, electrons and positrons do not cancel
in three-wave interactions. In deriving the three-wave probability the 
Alfv\'en waves are assumed to be linearly polarized with
$\be=(\cos\phi,\sin\phi,0)$ and $\be'=(\cos\phi',\sin\phi',0)$.
The wave vectors are written as $\bk=(k_\perp\cos\phi,k_\perp\sin\phi,k_\parallel)$
and $\bk'=(k'_\perp\cos\phi',k'_\perp\sin\phi',k'_\parallel)$. 
In the strong magnetic field, one has $R_S
\approx {k''}^2v^2_s/(2\omega^2_{pi})$ with $\omega_{pi}$ the ion plasma frequency. 
Using (\ref{eq:a}), one obtains
\begin{eqnarray}
u_{AA'S}(\bk,\bk',\bk'')&\approx&
2\pi r_ec^2\,{\hbar\omega\over m_ec^2}
\left({|k''_\parallel| c\over\omega'\beta_s}\right)\left({\omega_{pi}\over
\Omega_e}\right)^2\nonumber\\
&&\times\sin^2(\phi'-\phi),
\end{eqnarray}
where $r_e\approx 2.8\times10^{-15}\,\rm m$ is the classical electron radius.
Here we assume that the relevant waves are linearly polarized. For the 
three-wave probability to be nonzero, the two linear polarizations must be noncoplanar
($\phi'\neq\phi$). Denoting the left-hand side of (\ref{eq:Gam1}) by $\Gamma_{A'}$ 
one obtains
\begin{eqnarray}
\Gamma_{A'}&\approx& 4r_ec\,{\hbar\omega'\over m_ec^2}
\left({\omega_{pi}\over\Omega_e}\right)^2\nonumber\\
&&\times\!\int\! dk_\perp\,k_\perp 
{\partial\over\partial k_\parallel}\Bigl[k_\parallel 
N_A(k_\perp,k_\parallel)\Bigr]\Biggl|_{k_\parallel=-k'_\parallel},
\label{eq:Gam2}
\end{eqnarray}
where $N_A(\bk)=N_A(k_\perp,k_\parallel)$ is assumed to be axially 
symmetric. The growth rate is sensitive to the spectrum of the incoming Alfv\'en waves,
a result similar to that derived by \cite{cd74} in the MHD approximation in the
special case where the Alfv\'en waves are circularly polarized propagating
parallel or antiparallel to the magnetic field.

The propagation of Alfv\'en waves emitted from the surface due to coupling of the shear
waves is oblique $k_\perp\gg k_\parallel$ (Sec. 2), implying $N_A$ is anisotropic
in the $\bk$ space. As an example, we  consider
\begin{equation}
N_A(k_\perp,k_\parallel)\propto k^{\alpha-2}_\parallel\,\delta(k_\perp-\eta k_\parallel),
\label{eq:NA}
\end{equation}
where the parameter $\eta$ characterizes the anisotropy, which can be taken 
as $\eta\sim c/V_{sh}\gg1$ for the emission mechanism discussed in Sec. 2.
This corresponds to an energy spectrum $E_A(\omega)\sim \omega^\alpha$
of the incoming wave. It can be shown that the positive growth 
(corresponding to decay of the incoming waves) requires a spectrum
with a positive slope. This condition can be understood as follows. For a transition 
$A\to A'+S$, there is an inverse process $A'\to A+S$ that removes the $A'$ waves. 
For $A\to A'+S$ to be dominant over $A'\to A+S$, an inversion (i.e. a positive slope) in 
the occupation number $N_A$ is required. 
Assuming that the spectrum is peaked at $\omega\sim\omega_0$
with the positive slope characterized by a power index $\alpha>0$,
the growth rate is derived as
\begin{equation}
{\Gamma_{A'}\over \omega'}\approx {\alpha(\alpha+1)\over4\pi^2}
\left({\omega_{pi}\over\omega_0}\right)^2\,
\left({\delta B_A\over B}\right)^2.
\label{eq:Gam3}
\end{equation}
Although the parameter $\beta_s$ is not 
present explicitly in the growth rate, the condition $\beta_s\ll1$ is needed so that 
the three-wave resonance condition can be satisfied \citep{cd74}.

\begin{figure}
\includegraphics[width=7.5cm]{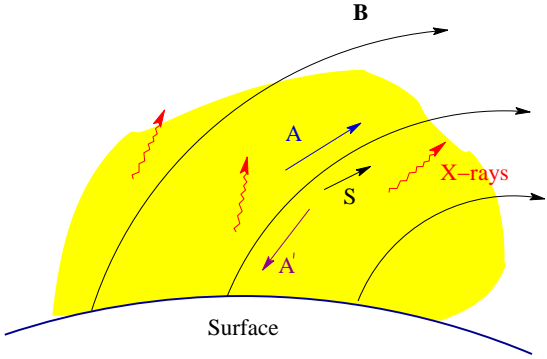}
\caption{Plasma heating due to decay of Alfv\'en waves near the surface.
The solid arrows indicate the group velocity of the relevant waves, which
are directed either parallel or antiparallel to the magnetic field direction.
The backward propagating wave may be reflected from the surface. So, the 
waves may be trapped near the star's surface. The ion sound wave
is assumed to be absorbed instantaneously leading to plasma heating. The shade represents 
a hot plasma (proton+electron) cloud emitting X-rays.}
\label{fig:trap}
\end{figure}

\section{X-ray emission from plasma heating}

In the following we specifically consider SGRs, which are characterized by
sporadic outbursts as well as occasional giant bursts. 
The burst emission from SGRs is generally considered to be magnetospheric origin, while
the persistent, pulsed emission may be attributed to surface emission.
Normal outbursts typically last weeks or months and 
consist of many ordinary, repetitive short bursts that last about 0.1 s,
which sets a lower limit to the damping time for the incoming Alfv\'en wave.
This damping time can be determined from the growth time of the 
backward propagating wave (cf. Eq. \ref{eq:Gam3}).

To estimate the growth rate one needs to assume a specific form of the 
incoming wave spectrum which should be determined by the shear waves in the crust. 
So far, no detailed model is available for the calculation of the latter. However, 
one may make an analogy with the energy spectra of shear waves in
earthquakes by assuming a spectrum peaked at a characteristic frequency
given by $\omega_0$ \citep{betal89}. The decay process discussed in Sec 3 occurs 
for $\omega'<\omega_0$. We assume that emission of Alfv\'en waves from the surface 
is accompanied by injection of plasmas of protons plus electrons (or/and positrons)
into the magnetosphere. The plasma density 
is not well constrained but its minimum density should be about the GJ density
$n_{GJ}={4\pi\varepsilon_0 B/eP}\approx (1.4\times10^{19}\,{\rm m}^{-3})
(P/5\,{\rm s})^{-1}(B/10^{11}\,{\rm T})$, where we ignore the cosine angle between the 
magnetic field and the rotation axis. 
Assuming $n_p\approx n_{GJ}\sim 10^{19}\,{\rm m}^{-3}$, $k'_\perp=300k_\parallel$. 
about the GJ density, $\alpha=2$, and $\delta B_A/B=5\times10^{-5}$, the growth time is estimated to be
$1/\Gamma_{A'}\sim 10^{-4}\,{\rm s}$ for $\omega_0\sim10^4\,{\rm s}^{-1}$.
This time is much shorter than the cascade time through $A\leftrightarrow A'+A''$
\citep{lm06}, implying that the decay occurs at a much faster rate than
pure Alfv\'enic cascades. The backward propagating wave generated from the process
may be reflected from the surface and propagates forward undergoing further decay. 
As a result, the waves are trapped close to the surface 
(at a distance $\Delta s=c/\Gamma_{A'}\approx 2\times10^4\,\rm m$ from the surface) 
without propagating to the opposite foot 
points (as shown in figure \ref{fig:trap}). The trapped waves 
continue to decay into ion sound waves which are absorbed by thermal
electrons. As a result, the plasma can be rapidly heated leading
to X-ray emission. Since the fraction of the wave energy dissipated in each
interaction is $\omega''/\omega\sim v_s/c\sim 10^{-3}$, the time for the trapped
waves to be absorbed is $(c/v_s)/\Gamma_{A'}\sim 0.1\,\rm s$,
which is comparable to the typical time scale of individual bursts. 

An attractive aspect of the mechanism discussed here is that
the X-ray emission from the decay of Alfv\'en waves is intrinsically variable on 
the time $1/\Gamma_{A'}$. Such time scale may be relevant for variabilities observed
in giant flares \citep{ws06}. However, to assess its relevance a quantitative model 
for the emission region is needed, which is beyond the scope of this paper and  
will be discussed elsewhere.

\section{Conclusions and discussion}

Transport of magnetic fields in the neutron star causes magnetic dissipation
both in the core and outside the star. The former leads to heating which may be radiated
through neutrino emission or partially conducts to the surface producing thermal
X-ray emission. In the latter case, the magnetic energy is efficiently transported to the 
magnetar's magnetosphere in Alfv\'en waves. These waves decay through
three-wave interactions. It is suggested here that because of the low 
$\beta_s\ll1$, the process involving a Alfv\'en wave decaying into ion sound waves that 
are then damped is the most efficient way to dissipate the magnetic energy.
In each interaction, the Alfv\'en wave cascades toward the small wavenumber regime.
Since the process is intrinsically dissipative converting the 
energy in the wave to kinetic energy in the plasma, such `inverse cascade' 
can be understood in analogy with the Langmuir condensation due to
nonlinear Landau damping which tends to pump the Lanmguir wave from 
larger to smaller wavenumber \citep{m86}. 
The growth rate for the scattered Alfv\'en wave, which quantifies
the efficiency of the process, is calculated in the random phase approximation
that is applicable provided that the waves generated have a broad 
range of frequencies and wave numbers. This approximation is relevant since the 
crust dislocations or fractures may produce many shear waves that are not related in 
phase (i.e. incoherent waves). It is shown here that the absorption time 
can be comparable to or shorter than the typical duration of individual outbursts, 
suggesting that such process can be important for transient X-ray emission.

Note that the decay process considered here was widely discussed in the MHD
approximation in the context of modulational instability or parametric decay with 
the application to the solar wind or solar flares \citep{ld76,g78,d78}. 
There were also recent discussions of this process using numerical
simulation \citep{zetal01}. These discussions mostly concentrated on 
circularly polarized Alfv\'en waves and assumed that the incoming wave is monochromatic
with the growth rate estimated to be $\Gamma_{A'}/\omega_0\sim(\delta B_A/B)/\beta^{1/2}_s$ in 
the limit $\beta_s=v_s/v_A\ll1$ \citep{go63,g78,d78}.  It is debatable whether 
such result is relevant for the solar wind and solar flares
because in both cases, the incoming waves should be broad band and 
it may not be appropriate to treat the incoming waves as a single coherent 
wave. Furthermore, in both cases the parameter $\beta_s=v_s/v_A$ is moderate.
At a relatively large $\beta_s$ the waves concerned (the relevant side-band waves) 
are away from the corresponding natural modes and as a result, the growth rate is 
considerably reduced. It is shown here that the decay process is favored in 
magnetars and can be an effective mechanism for magnetic dissipation required 
for the interpretation for the observed transient X-ray emission.   

\section*{Acknowledgement}
QL thanks Don Melrose for helpful discussions and comments.


\begin{thebibliography}{22}
\bibitem[\protect\citeauthoryear{Bhattacharjee \& Ng}{2001}]{bn01}
Bhattacharjee, A., Ng, C.S. Random scattering and anisotropic turbulence of shear 
Alfv\'en wave packets. ApJ 548, 318-322, 2001.
\bibitem[\protect\citeauthoryear{Blaes, Blandford, Goldreich, \& Madau}{1989}]{betal89}
Blaes, O., Blandford, R., Goldreich, P., Madau, P. Neutron starquake models for gamma-ray
bursts, ApJ 343, 839-848, 1989.
\bibitem[\protect\citeauthoryear{Chandran}{2005}]{c05}
Chandran, B. Weak compressible magnetohydrodynamic turbulence in the solar corona,
Phys. Rev. Lett. 95, 265004, 2005.
\bibitem[\protect\citeauthoryear{Cohen \& Dewar}{1974}]{cd74}
Cohen, R.H., Dewar, R.L. On the backscatter instability of solar wind Alfv\'en waves, 
JGR 79, 4174-4178, 1974.
\bibitem[\protect\citeauthoryear{Cumming, Arras, Zweibel}{2004}]{cetal04}
Cumming, A., Arras, P., Zweibel, E. Magnetic field evolution in neutron star crusts due to the
Hall effect and ohmic decay, ApJ 609, 999-1017, 2004.
\bibitem[\protect\citeauthoryear{Del Zanna, Velli, \& Londrillo}{2001}]{zetal01}
Del Zanna, L., Velli, M., Londrillo, P. Parametric decay of circularly polarized Alfvén waves:
Multidimensional simulations in periodic and open domains. A\&A 367, 705-718, 2001.
\bibitem[\protect\citeauthoryear{Derby}{1978}]{d78}
Derby, N. F. Modulational instability of finite-amplitude, circularly polarized Alfven waves.
ApJ 224, 1013-1016, 1978.
\bibitem[\protect\citeauthoryear{Galeev \& Oraevskii}{1963}]{go63}
Galeev, A.A., Oraevskii, V.N. The stability of Alfv\'en waves. Sov. Phys. Dokl. 7, 988-993, 1963.
\bibitem[\protect\citeauthoryear{Galtier et al.}{2002}]{getal02}
Galti\'er, S., Nazarenko, S.V., Newell, A.C., Pouquet, A. Anisotropic turbulence of shear-Alfv\'en
waves. ApJ 564, L49-L52, 2002.
\bibitem[\protect\citeauthoryear{Goldstein}{1978}]{g78}
Goldstein, M.L. An instability of finite amplitude circularly polarized Alfven waves.
ApJ 219, 700-704, 1978.
\bibitem[\protect\citeauthoryear{Heyl \& Hernquist}{1999}]{hh99}
Heyl, J.S., Hernquist, L. Nonlinear QED effects in strong-field magnetohydrodynamics.
Phys. Rev. D 59, 045005, 1999.
\bibitem[\protect\citeauthoryear{Heyl \& Hernquist}{2005}]{hh05}
Heyl, J.S., Hernquist, L. A QED model for the origin of bursts from soft gamma repeaters and
anomalous X-ray pulsars. ApJ 618, 463-473, 2005.
\bibitem[\protect\citeauthoryear{Iroshnikov}{1964}]{i64}
Iroshnikov, P.S. Turbulence of a conducting fluid in a strong magnetic field.
Sov. Astron. J. 7, 566-571, 1964.
\bibitem[\protect\citeauthoryear{Kaspi}{2004}]{k04}
Kaspi, V. Soft $\Gamma$ Repeaters and Anomalous X-ray Pulsars: Together Forever,
in: Young neutron Stars and Their Environments, IAU Symposium Vol. 218, p. 231-238, 2004.
\bibitem[\protect\citeauthoryear{Kraichnan}{1965}]{k65}
Kraichnan, R.H. Inertial-range spectrum of hydromagnetic turbulence. Phys. Fluids 8, 1385-1387, 1965.
\bibitem[\protect\citeauthoryear{Lashmore-Davies}{1976}]{ld76}
Lashmore-Davies, C.N. Modulational instability of a finite amplitude Alfv\'en wave. 
Phys. Fluids 19, 587-589, 1976.
\bibitem[\protect\citeauthoryear{Lithwick \& Goldreich}{2003}]{lg03}
Lithwick, Y., Goldreich, P. Imbalanced weak magnetohydrodynamic turbulence.
ApJ 582, 1220-1240, 2003.
\bibitem[\protect\citeauthoryear{Luo \& Melrose}{2006}]{lm06}
Luo, Q., Melrose, D.B. Anisotropic weak turbulence of Alfv\'en waves in collisionless astrophysical
plasmas. MNRAS, 368, 1151-1158, 2006.
\bibitem[\protect\citeauthoryear{Melrose}{1986}]{m86}
Melrose, D.B. Instabilities in Space and Laboratory Plasmas. 
Cambridge University Press, 1986.
\bibitem[\protect\citeauthoryear{Miller}{1991}]{m91}
Miller, J.A. Magnetohydrodynamic turbulence dissipation and stochastic proton acceleration in solar
flares. ApJ 376, 342-354, 1991.
\bibitem[\protect\citeauthoryear{Shebalin et al}{1983}]{smm83}
Shebalin, J.V., Matthaeus, W.H., Montgomery, D. J. Anisotropy in MHD turbulence due to a mean
magnetic field. Plasma Phys.  29, 525-547, 1983.
\bibitem[\protect\citeauthoryear{Thomspon \& Blaes}{1998}]{tb98}
Thompson, C., Blaes, O. Magnetohydrodynamics in the extreme relativistic limit.
Phys. Rev. D 57, 3219-3234, 1998.
\bibitem[\protect\citeauthoryear{Thompson \& Duncan}{1995}]{td95}
Thompson, C., Duncan, R. The soft gamma repeaters as very strongly magnetized neutron stars - I.
Radiative mechanism for outbursts. MNRAS 275, 255-300, 1995.
\bibitem[\protect\citeauthoryear{Watts \& Strohmayer}{2006}]{ws06}
Watts, A.L., Strohmayer, T.E. High frequency oscillations during magnetar flares.
ApSS, in press, 2006. (astro-ph/0608476) 
\bibitem[\protect\citeauthoryear{Woods et al}{2006}]{wetal06}
Woods, P.M., Kouveliotou, C., Finger, M.H., G\"og\"uş, E., Wilson, C.A., Patel, S.K., 
Hurley, K., Swank, J.H. The Prelude to and Aftermath of the Giant Flare of 2004 December 27:
Persistent and Pulsed X-Ray Properties of SGR 1806-20 from 1993 to 2005. ApJ 654, 470-486, 2006.
\end{thebibliography}
\end{document}